\documentclass[twocolumn,aps,prd]{revtex4}
\begin{document}
\title{
AdS/CFT correspondence in a Friedmann-Lemaitre-Robertson-Walker brane
}
\author{Takahiro Tanaka} 
\email{tama@scphys.kyoto-u.ac.jp}

\affiliation{
Department of Physics, Kyoto University, Kyoto 606-8502, Japan 
}
\begin{abstract}
According to the AdS/CFT correspondence conjecture,
the Randall-Sundrum infinite braneworld is equivalent to 
four dimensional Einstein gravity with ${\cal N}=4$ super Yang-Mills 
fields at low energies. 
Here we derive a four dimensional effective equation
of motion for tensor-type perturbations 
in two different pictures, and demonstrate their equivalence.
\end{abstract}
\maketitle
Braneworld scenarios proposed by Randall and
Sundrum\cite{Randall:1999ee, Randall:1999vf} 
attracted much attention. 
Especially the second model\cite{Randall:1999vf} (RS II) has been investigated 
a lot as a model which realizes
a new scheme of compactifying an extra 
dimension\cite{Langlois:2002bb,Maartens:2003tw}. 
In this model bulk dynamics is governed by 
the five dimensional Einstein equations with a negative cosmological constant, 
and ordinary matter fields are 
confined to a four dimensional brane 
located at a boundary of the bulk with $Z_2$-symmetry. 
The simplest unperturbed background is given by a five dimensional AdS bulk 
with a cosmological constant, 
\begin{eqnarray*}
ds^2=-{\ell\over z^2}\left(
     -dt^2 + \delta_{ij}dx^i dx^j + dz^2\right),
\end{eqnarray*} 
where $\ell$ is the curvature length of the AdS space. 
Latin indices are used for 3-dimensional
spatial coordinates and are raised and lowered by using 
the Kronecker delta $\delta_{ij}$.  
In the original RS II
model \cite{Randall:1999vf}, a Minkowski brane placed at a fixed
value of $z$ was considered. This configuration becomes a 
solution by tuning the tension of the brane as 
$\sigma=3/4\pi G\ell^2$, where $G$ is Newton's constant.  
Soon the model was extended so as to realize 
a general expansion law of the universe on the brane 
\cite{Binetruy:1999ut, Binetruy:1999hy, Mukohyama:1999qx, Ida:1999ui, 
Kraus:1999it}. In fact, once we introduce matter fields or detuned 
brane tension, the brane in general starts to move. 
As an easiest example, 
a moving brane in $z$-direction represents a flat 
Friedmann-Lemaitre-Robertson-Walker (FLRW) universe. 
The scale factor on the brane is given by $a=\ell/z(t)$, and 
the Hubble parameter $H$ is related to the brane motion as 
$H=-\dot z/\ell\sqrt{1-\dot z^2}$, where dot represents differentiation 
with respect to $t$. From the junction condition on the brane,  
a modified Friedmann equation 
\begin{equation}
{\cal H}^2={8\pi G\over 3}a^2\left(\rho+ {\rho^2\over 2\sigma}\right)
\label{FRW2}
\end{equation}
follows,  
where ${\cal H}\equiv aH$ and $\rho$ is the total energy density 
of matter fields localized on the brane. 

In the RS II model 
the relative correction to Newton's law at a distance $r$ 
is suppressed by a factor $\ell^2/r^2$\cite{GarTan,Giddings:2000}. 
Experiments of gravitational forces constrain 
$\ell$ to be shorter than $0.1$mm or so. 
The fundamental mass scale of 
five dimensional gravity, $m_5\equiv G_5^{-1/3}$, 
is related to the four dimensional 
Planck mass $m_{pl}\equiv 1/\sqrt{G}$ 
by $m_{5}^3=m_{pl}^2/\ell$. Hence $m_5$ must be  
larger than $10^8$GeV, which is far beyond the energy scale that 
collider experiments can reach. Therefore the most stringent constraint 
on models of this type is expected to be brought 
by examining the history of the early universe or 
by testing modification of gravitational forces. 

Aiming at providing constraints from 
the cosmic microwave background radiation, 
many works on cosmological perturbations in RS II model 
have been done\cite{Maartens:1999hf,Mukohyama:2000ui,Kodama:2000fa,Langlois:2000ia,Langlois:2000ph,perturbation_scalar,Koyama:2001rf,Langlois:2000ns,Gorbunov:2001ge,Frolov:2002qm,Kobayashi:2003cn,Hiramatsu:2003iz,Easther:2003re,Battye:2003ks,Ichiki:2003hf}.
However, we have not clearly understood yet how to qualitatively 
estimate the leading order correction to the predictions 
of the standard cosmology. 
First of all, once we take into account the effect of 
higher dimensions, solving perturbation equations 
for given initial conditions is difficult even at the classical level 
since in general at least two dimensional 
field equation must be solved. Moreover, it is not very clear 
which initial conditions are appropriate in principle in the 
braneworld setup. One exceptional case is creation of 
a braneworld from nothing\cite{Garriga:1999bq}. In this case 
the standard scheme for specifying boundary conditions for 
the wavefunction of the universe (no boundary or tunneling 
conditions) will work. 
Even if we restrict our consideration 
to such a well-posed setup, it is still a very tough problem 
to sum up the fluctuations from all independent Kaluza-Klein 
modes\cite{Gorbunov:2001ge,Kobayashi:2003cn}. 
Unless a de Sitter brane is concerned, decomposition of 
independent modes is quite non-trivial. Even if we succeed in 
solving all independent mode functions, there is a problem of 
ultra-violet divergence when we evaluate the value of a bulk 
field exactly on the brane\cite{Kobayashi:2000yh,Sago:2001gi}. 
This problem has not been taken seriously in literature 
since the divergence is logarithmic for 
the expectation value of a squared bulk field. 
However, if we compute quantities which contains more 
differentiations of a field, divergence will become severer. 

On the other hand, non-linear corrections to
gravity on a Minkowski brane were also
investigated\cite{GiaRen,KudTan,Wiseman}.
In all cases examined so far, 
corrections are always suppressed by a factor $\ell^2/r^2$.
Namely, any non-linear effects do not make it easier to 
discriminate braneworld models of the RS II type from 
the standard. However, we proposed a conjecture that 
there is no stationary large black hole solution in the RS II
model\cite{Tanaka2002, Emparan:2002px}. 
The conjecture indicates that a black hole localized on the brane 
evaporate even at the classical level. 
Some numerical works to construct a large black hole solution were 
done recently, but only small black holes were found\cite{Kudoh1,
Kudoh2}. (See also \cite{Bruni:2001fd,
Karasik:2004wk,Casadio:2004nz}.) 
This fact might be a supporting evidence for the conjecture, 
although different interpretation is also possible. 
If the conjecture is correct, the life time of a black hole 
is estimated as $\tau\approx (M/M_\odot)^3(1mm/\ell)^2\times 1$ year, 
where $M$ and $M_{\odot}$ are masses of a black hole and the sun, 
respectively.  
The above conjecture is based on the AdS/CFT 
correspondence\cite{Maldacena,HawHer}.  
The AdS/CFT correspondence conjecture asserts 
that the effective action 
of ${\cal N}=4$ super Yang-Mills fields 
evaluated on the metric induced on the boundary 
is given by 
\begin{eqnarray}
 W_{CFT}=S_{EH}+S_{GH}-S_1-S_2-S_3, 
\label{AdSCFT}
\end{eqnarray}
where $S_{EH}=-{1\over 16\pi G_5}\int d^5x\sqrt{-g}\left(
 ^{(5)}R+{12\over \ell^2}\right)$, 
$S_{GH}=-{1\over 8\pi G_5} \int d^4x \sqrt{-^{(4)}\! g}\,K$, 
$S_1 = -(3/8\pi G_5 \ell) \int d^4 x \sqrt{-^{(4)}\! g}$, 
$\left.S_2 = -(\ell/ 32\pi G_5) \int d^4 x \sqrt{-^{(4)}\! g}\,^{(4)}\! R ~
\rm{and}~ S_3=-(\ell^3/64\pi G_5)\right.$ 
$ \ln(z_0/z)\int d^4 x \sqrt{-^{(4)}\! g}
\left(^{(4)}\! R_{\mu\nu}\,^{(4)}\! R^{\mu\nu}-(1/3)\,^{(4)}\! R^2\right)$.
$z_0$ determines the renormalization scale. 
$K$ is the trace of the extrinsic curvature of the boundary. 
Left hand side is the action for five dimensional gravity theory. The counter 
terms $S_1$, $S_2$ and $S_3$ are necessary to cancel manifest dependencies 
on the boundary location. 
Using this equality, the action of the RS II model is rewritten as 
\begin{eqnarray}
 S_{RS} & = & 2\left(S_{EH}+S_{GH}\right)
    -2S_1 +S_{matt}\cr
     &  = &2 S_2+2\left(W_{CFT}+S_3\right)+S_{matt}. 
\label{AdSCFT2}
\end{eqnarray}
We notice that $2 S_{2} $
is the ordinary four dimensional Einstein-Hilbert action,  
while $W_{CFT}+S_3$ is the effective action for a conformal field 
theory (CFT) with an ultra-violet cutoff. 
The above formula indicates equivalence between the RS II model 
and four dimensional Einstein gravity with CFT. 
Here we note that the leading order corrections due to 
the bulk effect at the classical level in the five dimensional RS picture 
come from one-loop quantum effects of CFT in the four dimensional picture. 
So far direct and satisfactory 
confirmations of the equivalence between these two pictures 
are limited to the following two cases. 
One is the case of linear perturbations from 
Minkowski brane\cite{DufLiu}, and the other is 
the homogeneous cosmology\cite{ShiIda}. 
Equivalence is satisfied up to $O(\ell^2)$. 

The main purpose of this letter is to add another example for the 
equivalence between RS II model and four dimensional Einstein gravity with CFT 
by considering tensor-type 
perturbations on a FLRW brane. We derive the 
leading order corrections at low energies in both 
four and five dimensional pictures, and show that in fact they are identical.  
Although here in this short article 
we shall not pursue an application of our new results to 
understand cosmological perturbations in the braneworld, 
we expect that it might open up a new 
approach to this problem. In the four dimensional picture we just need 
to consider the backreaction due to vacuum polarizations 
of CFT, and there are only two graviton degrees of 
freedom for each three dimensional wave number. This fact 
may allow us to avoid the problems mentioned above 
in solving cosmological perturbations. 

{\it Five dimensional RS picture}: 
First we discuss 
tensor-type perturbations on a FLRW brane 
in the five dimensional RS picture. 
The bulk metric is given by 
\begin{equation}
 ds^2={\ell^2\over z^2}\left(
     -dt^2 + (\delta_{ij}+h_{ij})dx^i dx^j + dz^2\right).  
\end{equation}
By using $Y_k^{ij}$, a transverse traceless tensor harmonics 
normalized as 
$
\int d^3 x\, Y_{k\,ij}({\bf x}) Y_{k'}^{ij}({\bf x})=\delta^3
 ({\bf k-k'}),  
$
we expand perturbations as 
$h_{ij}=Y_{k\,ij} \Phi$.
In these coordinates the five dimensional perturbation equation 
reduces to 
\begin{equation}
 \left(-\partial_z^2 +{3\over z}\partial_z+\partial^2_t+k^2\right)
 \Phi=0.
\label{5Deq}
\end{equation}
The general solution to this equation is given by 
\begin{eqnarray}
 \Phi & = & \int d\omega\, \tilde\Psi(\omega) e^{-i\omega t} 2(pz)^2
     K_2(pz)\cr
    & = & \int d\omega\, \tilde\Psi(\omega) e^{-i\omega t} 
\cr && \quad \times
    \left[1-{(pz)^2\over 4} 
           +{(pz)^4\over 16}
              \left(b-\ln(pz)\right)+\cdots\right],  
\end{eqnarray}
where 
$p^2= -(\omega+i\epsilon)^2+k^2$, 
$b={1\over 2}\left({3\over 2}-2\gamma\right)+\ln 2$, and 
$\gamma$ is Euler's constant. 
Here we have chosen the branch cut of the modified Bessel function $K_2$ 
so that there is no incoming wave from past null infinity in the bulk. 
What we have to do is to choose $\tilde\Psi$ such that 
$\Phi$ satisfies the perturbed junction condition
$n^\rho\partial_\rho\Phi=0$, 
where $n^\rho$ is the unit normal to the brane. 
Notice that the conformal time of the 
induced metric on the brane 
$\eta$ is related to the bulk conformal time $t$ 
by $d\eta=\sqrt{1-(dz(t)/dt)^2}dt$. 
The derivative along the direction parallel to the brane 
is given by  
$
 \partial_\eta=\sqrt{1+(H\ell)^2}\partial_t-H\ell \partial_z.
$
Thus the unit normal is given by 
$n^\mu\partial_\mu=a^{-1}(-H\ell\partial_t
+\sqrt{1+H^2\ell^2}\partial_z)$. 

Here we use an alternative approach. 
In Ref.~\cite{Shiromizu:1999wj} effective Einstein equations
\begin{equation}
 ^{(4)}G^{\mu}_{~\nu}-8\phi G_4 T^{\mu}_{~\nu}
   =(8\pi G_5)^2\pi^{\mu}_{~\nu} -E^\mu_{~\nu},
\label{SMSeq}
\end{equation}
were derived, where $\pi^{\mu}_{~\nu}$ is a tensor quadratic in 
the energy momentum tensor
$T^\mu_{~\nu}$ and $E^{\mu}_{~\nu}$ is a projected Weyl tensor 
defined by $n^\rho n^\sigma C^\mu_{~\rho\nu\sigma}$. 
From Eq.~(\ref{SMSeq})
the effective four dimensional equation for tensor perturbations 
is given by 
\begin{equation}
 \left(\partial_\eta^2+2{\cal H}\partial_\eta+k^2 \right)\phi
 =-2E, 
\label{phieq}
\end{equation}
with $\phi\equiv \Phi|_{z=z(t)}$, and  
the correction due to $E_{\mu\nu}$ is explicitly given as 
\begin{eqnarray}
 -2E &=& \Biggl\{
       (H\ell)^2\left(\partial_t^2 + \partial_z^2 \right)
       -2 H\ell\sqrt{1+(H\ell)^2}\partial_t \partial_z 
\cr &&\quad   
       +\left(\partial_z^2-{1\over z}\partial_z\right)
      \left.\Biggr\}\Phi\right|_{z=z(t)}, 
\end{eqnarray}
To Solve Eq.~(\ref{5Deq}) supplemented with Eq.~(\ref{phieq}) 
is equivalent 
to solve the same equation with the perturbed junction condition, 
$n^\rho\partial_\rho\Phi=0$. To see deviation from the four 
dimensional Einstein gravity, the former is more convenient. 

At low energies (when $H^2\ell^2\ll 1$) 
further reduction is possible in an approximate sense like 
\begin{eqnarray}
 && \partial_t^2\Phi\approx \partial_\eta^2 \Phi,\cr
 && \partial_z^2\Phi
    \approx -{1\over 2}\int d\omega\, \tilde\Psi e^{-i\omega \eta} p^2
    \approx -{1\over 2} \left(\partial_\eta^2 +k^2\right)
               \Phi,\cr
 && \partial_t\partial_z\Phi
    \approx -{\ell \over 2a} \partial_\eta 
           \left(\partial_\eta^2 +k^2\right)\Phi. 
\end{eqnarray}
Here we have neglected higher order corrections of $O(\ell^4)$. 
Only the last term $\left(\partial_z^2-{1\over z}\partial_z\right)\Phi$ 
does not allow such a simple reduction because the third term 
in the expansion of $K_2(pz)$, which is not a polynomial in $p^2$, 
contributes to this term. 
This term does not have explicit $\ell^2$ suppression at this 
level. However, first two terms in the expansion of $K_2(pz)$ 
vanish for this combination of differentiation. 
As a result, a factor of $z^2=\ell^2/a^2$ arises. 
We finally obtain 
\begin{eqnarray}
 &&\hspace{-10mm}\left(\partial_\eta^2+2{\cal H}\partial_\eta+k^2 \right)\phi
\cr
  &&\approx  \left[
     {(H\ell)^2\over 2}\left(\partial_\eta^2 -k^2 \right)
     +{H\ell^2\over a}\partial_\eta \left(\partial_\eta^2 +k^2 \right)
    \right]\phi\cr
   &&~+{\ell^2\over 2a^2}\int d\omega\, \tilde\phi e^{-i\omega\eta} 
             p^4 \left(b-{3\over 4}-\ln\left[{p \ell\over a}\right]\right). 
\label{RSEQ}
\end{eqnarray}
A similar equation was derived for scalar-type perturbations 
in Ref.~\cite{perturbation_scalar}. 
All the corrections are suppressed by $\ell^2$ or $\ell^2\ln \ell$. 
The first term on the right hand side can be rewritten 
by using the lower order equation as 
\begin{eqnarray*}
    {\ell^2 \over a^2}\left[
     \left(3{\cal H}^3-2{\cal H H}'\right)\partial_\eta
       +k^2{\cal H}^2
    \right]\phi.
\end{eqnarray*}

{\it Four dimensional CFT picture}: 
Effective equations of motion for tensor perturbations 
in CFT picture have already been obtained in Ref.~\cite{staro}. 
Here we give a brief derivation of it. 
The equation for the tensor-type perturbation 
is given in the form of  
\begin{equation}
 \partial_\eta a^2 \partial_\eta  \phi
  +a^2 k^2  \phi- 16\pi G a^2\tau =0, 
\end{equation}
where the perturbation variable $\phi$ is defined 
as before, and 
$
 \tau(\eta)\equiv \int d^3 x \, T^{(CFT)}_{ij}(\eta,{\bf x}) Y_k^{ij}({\bf x})
$
is the contribution from the 
effective energy momentum tensor 
of CFT. 
In order to evaluate the energy momentum tensor due 
to vacuum polarization of CFT, we can make use of the fact 
that the metric of flat perturbed FLRW universe 
is related via conformal transformation 
to Minkowski spacetime with the corresponding perturbations as 
\begin{equation}
 ds_{(1)}^2=g_{\mu\nu}dx^\mu\, dx^\nu
      =a^2(\eta)ds_{(0)}^2, 
\end{equation}
where 
$
 ds_{(0)}^2=g_{\mu\nu}^{(0)}dx^\mu\, dx^\nu=
      \left(\eta_{\mu\nu}+ h_{\mu\nu}\right)dx^\mu\,dx^\nu
$. 
We consider one parameter family of conformally related metrics 
$g^{(\theta)}_{\mu\nu} = a^{2\theta} g^{(0)}_{\mu\nu}$ 
connecting $g_{\mu\nu}\equiv g^{(1)}_{\mu\nu}$ with $g^{(0)}_{\mu\nu}$. 
The action for CFT is invariant under conformal 
transformation except for $T^{(A)}_{\mu\nu}$,  
the contribution from the conformal anomaly.  
Hence, the energy momentum tensor of CFT excluding the 
anomaly contribution 
transforms in a trivial manner under a conformal transformation. 
Thus we have  
\begin{equation}
  T^{(CFT)}_{\mu\nu}=a^{-2} T^{(0)}_{\mu\nu}+T^{(A)}_{\mu\nu}, 
\end{equation}
where $T_{\mu\nu}^{(0)}$ is the CFT energy momentum tensor evaluated on 
the metric $g_{\mu\nu}^{(0)}$. 
Correspondingly, we divide $\tau$ into two pieces,  
$\tau^{(0)}$ and  $\tau^{(A)}$.  

From the comparison of Eqs.~(10), (12) and (13) in Ref.\cite{DufLiu}, 
$\tau^{(0)}$ is found to be given by 
\begin{eqnarray}
  -16\pi G a^2 \tau^{(0)} 
        &= & 2 \int d\omega\, e^{-i\omega\eta} 
           \Pi_2 \!\left(p^2\right)p^4
            \tilde \phi(\omega),  
\end{eqnarray}
with 
\begin{equation}
  \Pi_2(p^2)={\ell^2\over 8}
         \ln{p^2\over\mu^2}, 
\end{equation}
where we used the properties of tensor-type perturbations, 
$h^{0 \alpha}=0$, $h^{~j}_j=0$, 
and $h^{ij} k_j=0$. 
Constant part in $\Pi_2$ was absorbed by the renormalization scale $\mu$. 
Here $\tilde \phi(\omega)\equiv (1/2\pi)\int d\eta\,
             e^{i\omega\eta} \phi(\eta)$ 
is the Fourier transform 
of $ \phi(\eta)$. 

Next we consider the anomaly contribution $\tau^{(A)}$. 
The anomaly contribution to the effective action, $S^{(A)}$, is 
given by an integral of the Seeley-deWitt 
coefficient as\cite{BD,HawHer} 
\begin{equation}
 S^{(A)}
        =-{1\over 16\pi^2}\int_0^1 d\theta \int d^4x\, a_2^{\theta} 
     \ln a\, ,
\end{equation}
with 
$ a_2^{\theta} = {N^2\over 6}\sqrt{-g^{(\theta)}}
                    (3R_{\mu\nu}R^{\mu\nu}-R^2)[g^{(\theta)}]$, 
and $N^2=\pi\ell^2/G$ is the number of degrees of freedom of CFT. 
The anomaly contribution to the 
energy momentum tensor is obtained by taking 
variation of $S^{(A)}$ as
$T^{(A)}_{\mu\nu}=-{1\over 2}({\delta S^{(A)}/\delta g^{\mu\nu}})$.
Expanding the action up to second order in $ h_{\mu\nu}$, we obtain 
\begin{eqnarray}
 S^{(A)}&=&{\ell^2\over 64\pi G}
  \int d\eta  
    \Biggl(2{\cal H}^4 
\cr &&\quad +\sum_k
     \phi
   \left[\ln a\left(
    \partial_\eta^2+k^2\right)^2+ 
    {\cal O}\right] \phi\Biggr), 
\label{SA}
\end{eqnarray}
where
\begin{eqnarray}
{\cal O} &= & 2{\cal H}\partial_\eta^3
          +\left({\cal H}'+{\cal H}^2\right)\partial_\eta^2
          +\left(({\cal H}^2)'+{\cal H} k^2 \right)\partial_\eta
\cr &&\quad
          +k^2 \left({\cal H}'-{\cal H}^2\right) 
          +4{\cal H}'{\cal H}^2-{\cal H}^4.
\end{eqnarray}

The first term in the round brackets in Eq.~(\ref{SA}) 
gives a correction to the background Friedmann equation. 
Variation of the action with setting $\phi=0$ leads to 
\begin{equation}
 aa''= {4\pi G\over 3} a^4 (\rho-3 P)+{\ell^2\over 2}{\cal H}^2{\cal H}'. 
\label{FRW1}
\end{equation}
Using the continuity equation
$ \rho'=-3{\cal H}(\rho+P)$,
we can integrate Eq.~(\ref{FRW1}) once to obtain 
\begin{equation}
{\cal H}^2={8\pi G\over 3}a^2\rho+{\ell^2\over 4a^2}{\cal H}^4 +{C\over a^2}.
\end{equation}
Here $C$ arises as an integration constant. 
This term represents the so-called dark radiation. 
It is easy to verify the equivalence 
between Eqs.~(\ref{FRW2}) and (\ref{FRW1}) 
up to $O(\ell^2)$ when $C=0$.

The $\phi$-dependent part of the anomaly contribution is given by 
\begin{eqnarray}
 16\pi G a^2 \delta\tau^{(A)} & = &
   16\pi G {\delta S^{(A)}\over \delta  \phi}
\cr   & = & {\ell^2\over 2} \left[\ln a \left(
    \partial_\eta^2+k^2\right)^2 +{\cal O}\right] \phi.
\end{eqnarray}
Combining two contributions $\tau^{(0)}$ and $\delta\tau^{(A)}$, using 
the lower order equation, we can write down the modified 
equation of motion for $\phi$ as  
\begin{eqnarray}
&&\left(\partial_\eta^2+2{\cal H}\partial_\eta +k^2+ 
  (4{\cal H}'+2{\cal H}^2)+16\pi G a^2 P\right)\phi\cr
  &&\quad\approx {\ell^2 \over a^2}\left[
     \left(3{\cal H}^3-2{\cal H H}'\right)\partial_\eta
       +k^2{\cal H}^2+2{\cal H}^2{\cal H}'-{1\over 2}{\cal H}^4
    \right]\phi\cr
   &&\qquad
   -{\ell^2\over 2 a^2}\int d^4p\, \tilde \phi e^{-i\omega \eta} 
             p^4 \ln(p/a\mu). 
\end{eqnarray}
Using the background equation, we can rewrite the above equation as 
\begin{eqnarray}
&& \left(\partial_\eta^2+2{\cal H}\partial_\eta +k^2 
  \right)\phi\cr
  &&\quad\approx {\ell^2 \over a^2}\left[
     \left(3{\cal H}^3-2{\cal H H}'\right)\partial_\eta
       +k^2{\cal H}^2+{2C\over a^2}\right] \phi\cr
   &&\qquad
   -{\ell^2\over 2a^2}\int d^4p\, \tilde \phi e^{-i\omega \eta} 
             p^4 \ln\left({p\over a\mu}\right). 
\end{eqnarray}
Hence, we find that the above equation for tensor-type perturbations 
is identical to Eq.~(\ref{RSEQ}) 
obtained in the five dimensional RS II picture, 
as far as the dark radiation term, which we neglected in 
deriving Eq.~(\ref{RSEQ}), is set to zero. 

In the cosmological 
context basically we have two non-dimensional quantities of
$O(\ell^2)$. One is $\ell^2 H^2$ and the other is $\ell^2 k^2/a^2$. 
In this paper we developed a perturbative expansion scheme which 
is valid at low energies up to $O(\ell^2)$ since we are interested 
in the AdS/CFT correspondence. However, the method taken in the 
five dimensional RS II picture admit an extension to more general cases 
relaxing the constraint $\ell H^2\ll 1$. 
This extension will be discussed in our forthcoming paper.

The author would like to thank Roy Maartens, Shirnji Mukohyama 
and Tsutomu Kobayashi for valuable discussions. 
This work was supported in part by the Monbukagakusho Grant-in-Aid
Nos.~12740154 and 14047212 
and by the Inamori Foundation.


\end{document}